\documentclass[9pt,twocolumn,twoside]{opticajnl}
\journal{opticajournal} 

\setboolean{shortarticle}{true}

\usepackage{amsmath}
\usepackage{amssymb}
\usepackage{graphicx}
\usepackage{dcolumn}
\usepackage{bm}
\usepackage{soul}
\usepackage{float}

\usepackage{lineno}

\title{Measurements of extreme first passage times in photon transport}

\author[1,*]{Aileen N. Carroll-Godfrey}
\author[1]{Eric I. Corwin}

\affil[1]{Department of Physics and Materials Science Institute, University of Oregon, Eugene, Oregon 97403, USA.}
\affil[*]{aileeng@uoregon.edu}

\begin{abstract}
    Photon transport through turbid media has typically been modeled through diffusion or telegraph equations. These models describe behavior of the average, or typical, photon with remarkable accuracy, however, we show here that they fail to capture the Extreme First Passage Times (EFPTs) of photon transport. By sending ultra-fast bursts of photons through a scattering medium and timing the arrival of the first passage photon, we measure the distribution of these EFPTs of photons in a random environment. Our measured EFPTs differ from those predicted by both the diffusion approximation and telegraph equation. Instead, we observe the EFPT as the time expected for light to travel through an index-averaged medium. These results reveal flaws in both models and invite a re-examining of their underlying assumptions.
\end{abstract}

\setboolean{displaycopyright}{false} 

\begin{document}

\maketitle


\textbf{Introduction. }When light passes through a scattering medium, the spatial distribution of photons is smeared out by random scattering events, resulting in a corresponding broadening of the distribution of photons from a pulsed source. This broadening has historically been modeled at short times by the telegraph equation~\cite{goldstein_diffusion_1951,ishimaru_diffusion_1989,durian_photon_1997,lemieux_diffusing-light_1998} and at asymptotically long times by the diffusion approximation~\cite{haskell_boundary_1994,bohren_absorption_1983,ishimaru_wave_1997}.  However, these models both rely on a central assumption, that the scattering, and thus the path through the system, of each photon is independent and identically distributed.  Here, we probe the quality of these models by examining a measurement of the Extreme First Passage Time (EFPT), that is, the \textit{first} time of first arrival within a pulse of light.  We use a femtosecond laser to fire pulses of light through a tunable scattering medium and measure the EFPT with a Single Photon Avalanche Diode (SPAD).  We present experimental evidence that both of these models provide quantitatively and qualitatively wrong predictions for this measurement. 

In this Letter, we review the history of models for the first passage of photons, then describe an experimental setup to measure photon EFPT. We describe the diffusion approximation and telegraph equation approaches to modeling photon transport, and compare EFPT predictions from these models to experimental measurements. We demonstrate that both models fail to predict the EFPT.


\begin{figure}[ht]
\includegraphics[width=\columnwidth]{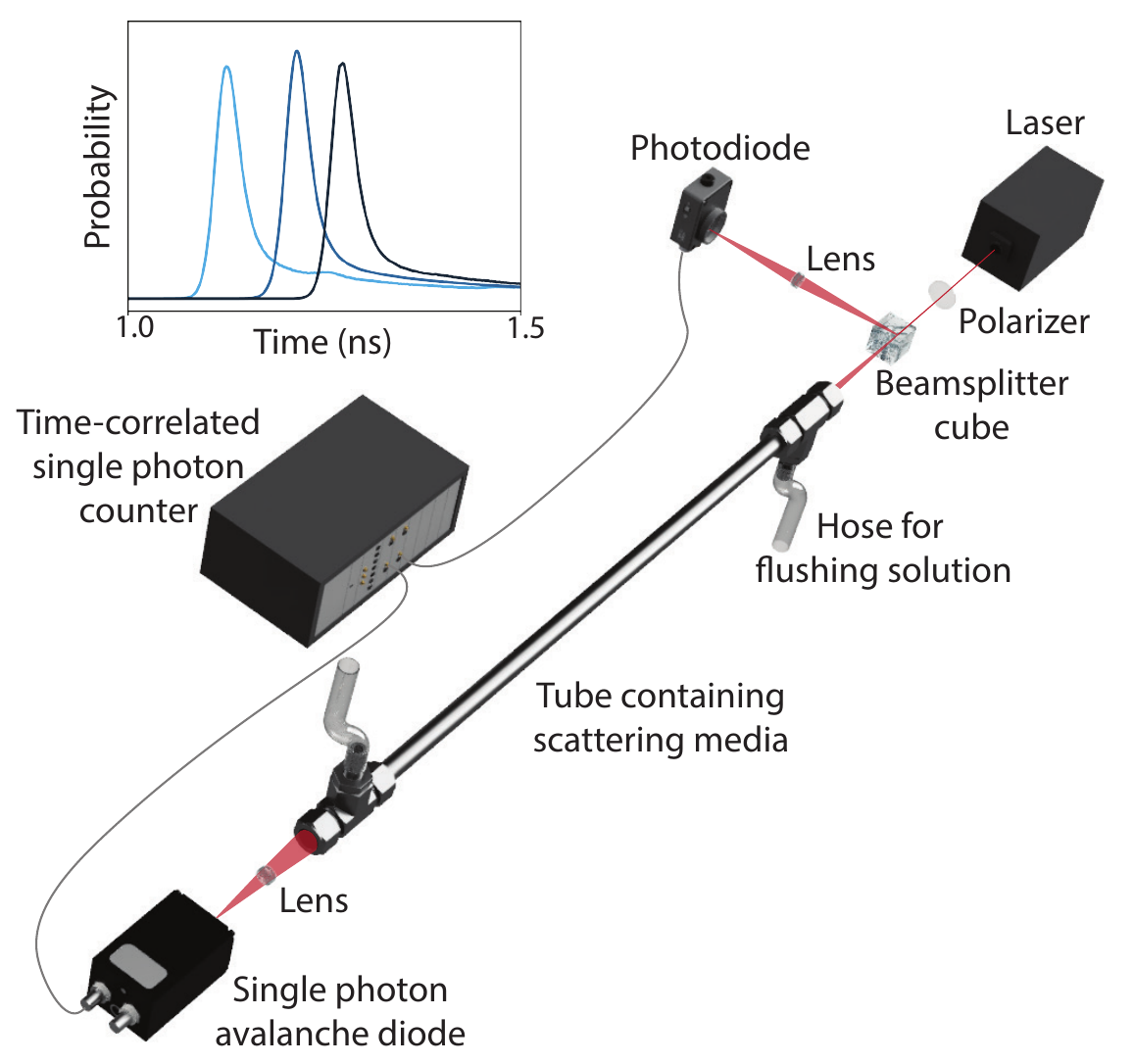}
\caption{\label{fig:setup}  Experimental setup, consisting of elements described in the text. Signals are collected over an adjustable time period, resulting in a histogram of EFPTs. An example of such a histogram at three different concentrations of scattering media is shown, with concentration increasing from left to right, where the x axis is the time relative to the speed of light in a vacuum, and the y axis is the probability of that time as the EFPT.}
\end{figure}

\textbf{Background. }Two widely used approximations for modeling photon transport are the diffusion approximation and the telegraph equation. The diffusion approximation predicts the bulk or average behavior of light pulses in random media and offers valuable insights into various systems across a broad range of applications~\cite{haskell_boundary_1994,  chandrasekhar_stochastic_1943,chandrasekhar_radiative_1950,  brown_light_1975, allgaier_diffuse_2021, taitelbaum_diagnosis_1999, amendola_accuracy_2021}. However, the diffusion approximation fails to incorporate ballistic photons and the speed of light, making it unsuitable for short length and time scales, media with significant anisotropy or absorption, or environments with low scattering~\cite{ishimaru_wave_1997, yoo_time-resolved_1990, yoo_when_1990, durian_photon_1997, zhang_wave_2002, pierrat_photon_2006}. The telegraph equation overcomes many of these limitations by accounting for the finite speed of light in a medium and capturing wave-like effects at small distances, providing a more accurate description of the bulk or typical behavior of light pulses~\cite{goldstein_diffusion_1951, durian_two-stream_1996, durian_photon_1997, lemieux_diffusing-light_1998, masoliver_solutions_1992, masoliver_finite-velocity_1996, das_non-fickian_1998, polishchuk_photon-density_1997}. Discrepancies remain at distances near the source relative to the transport mean free path~\cite{durian_photon_1997, dudko_photon_2005, lemieux_diffusing-light_1998}, and the accuracy of the predicted speed of light depends on how the scattering path length and anisotropy are treated during the derivation~\cite{ heizler_asymptotic_2012, hoenders_telegraphers_2005, polishchuk_photon-density_1997}.

First passage times have been broadly described and studied across various diffusive systems~\cite{redner_guide_2001,weiss_applications_2002,godec_first_2016,noskowicz_average_1988}, including chemical and molecular reactions~\cite{redner_guide_2001,grebenkov_molecular_2021}, protein interactions~\cite{polizzi_mean_2016}, fractal media~\cite{chun_heterogeneous_2023}, and stock market fluctuations~\cite{barney_first-passage-time_2017,zsurkis_first_2024}. In general, extreme value distributions, which describe values drawn from the tails of a source distribution, exhibit distinct statistical behaviors~\cite{lawley_distribution_2020,lawley_universal_2020} and play a critical role in processes such as menopause timing~\cite{lawley_slowest_2023}, gene activation~\cite{schuss_redundancy_2019}, and oocyte fertilization by sperm~\cite{meerson_mortality_2015,schuss_redundancy_2019}.  In photonic systems, changes in pulse shape due to absorption and scattering have been measured~\cite{lee_using_2007,madsen_experimental_1992,ishimaru_diffusion_1978,yoo_time-resolved_1990,yoo_when_1990,zhang_wave_2002,calba_ultrashort_2008}, and photon first passage time distributions have been characterized numerically and experimentally~\cite{rossetto_isotropic_2022,long_particle_2001,weiss_applications_2002,zeller_light_2020}. To our knowledge, EFPT distributions have not previously been experimentally measured.


\textbf{Experimental methods. }Our experimental setup is shown in Figure \ref{fig:setup}. Optical pulses of 140 fs in duration are generated with a femtosecond pulsed laser (Coherent Chameleon Ultra II laser) operating at central wavelengths tunable between 680-1080 nm with a repetition rate of 80 MHz and wavelength-dependent total output power range of 650-3500 mW. These pulses are sent through a power attenuating polarizer (Thorlabs GL10-B).  The majority of the power is bled off through a beamsplitter (Thorlabs BS041) and sent to a photodiode (Thorlabs DET10A).  This photodiode triggers a time-correlated single photon counter (TCSPC) (PicoQuant HydraHarp 400) to start a time-of-flight measurement.  The remainder of the light continues into a 12.7mm diameter, 1.015 m long stainless steel tube that has been internally polished.  This tube is filled with a solution of water and 20 nm diameter silica nanospheres (LUDOX AS-40). The nanospheres have an index of refraction $n$ = 1.453-1.456, dependent on wavelength.  The concentration of nanospheres is variable, in place, from volume fractions of 0 to 0.235 parts silica to 1 part water, where the maximum value is set by the volume fraction of undiluted LUDOX. The  concentration is adjusted by flushing the solution and flowing in a new solution.  After traveling through the scattering medium, photons exit the tube and are concentrated through a doublet lens (Thorlabs AC254-030-B-ML) onto a Single Photon Avalanche Diode (Micro Photon Devices PDM series SPAD).  The signal from the SPAD triggers the TCSPC to stop and the time of flight measurement is recorded. 

The TCSPC and SPAD both have a dead-time less than 80 ns, yielding an effective experimental repetition rate of 12.5 MHz. Using a detector for the TCSPC start and stop signals allows for measurement resolution that is not limited by the dead time. Integration times varied from 1-600 s.

The SPAD detection efficiency varies with wavelength from 10-30\%. From the peak laser power, approximately 3.5 W at 750 nm light, repetition rate of 80 MHz, and power attenuation from the optical components, we find a generous upper bound for the number of photons per pulse, $N$, entering the scattering medium as $5 \times 10^{6}$.

We use Ludox AS-40 as a scattering medium because it scatters strongly, absorbs minimally, is easily diluted with water, is slow to aggregate, and aggregation is easily reversed through sonication. Ludox has been well characterized for light scattering purposes~\cite{dezelic_determination_1960,bonnelycke_light_1959,goring_light-scattering_1957}. The small size of the silica nanospheres (roughly 3\% of the wavelengths used in these experiments) places us firmly within the Rayleigh scattering regime for which we can calculate $\sigma_{s}$ and $\mu_{s}$, the scattering cross section and coefficient, as~\cite{bohren_absorption_1983, chandrasekhar_radiative_1950} 
\begin{equation}\label{eq:scatter}
    \sigma_{s} =  \frac{2 \pi^{5}}{3} \frac{d^{6}}{\lambda^{4}}\left(\frac{n_{silica}^{2}-n_{water}^{2}}{n_{silica}^{2}+2n_{water}^{2}}\right)^{2},\ \mu_{s} = \frac{C}{\frac{4}{3} \pi \left(\frac{d}{2}\right)^{3}} \sigma_{s}
\end{equation}
where $d$ is the scatterer diameter, $\lambda$ is the wavelength of light being scattered, $n$ is the refractive index of the scatterer material, $C$ is the volume concentration of scatterers, and the scattering coefficient is the inverse of the scattering path length with units $\frac{1}{\textrm{m}}$. Using measurements performed with a UV/Vis/NIR spectrophotometer (Perkin Elmer Lambda-1050), we further characterize this medium by measuring $\sigma_{a}$, the absorption cross section, and use the Beer-Lambert law to find $\mu_{a}$, the absorption coefficient, as~\cite{lakowicz_principles_2006} 
\begin{equation}\label{eq:absorb}
    \sigma_{a} = 1.53\times 10^{-22}\ \textrm{m}^{2},\ \mu_{a} = \frac{C}{\frac{4}{3} \pi \left(\frac{d}{2}\right)^{3}} \sigma_{a}.
\end{equation}
%
%

\begin{figure}[hp]
\includegraphics[width=\columnwidth]{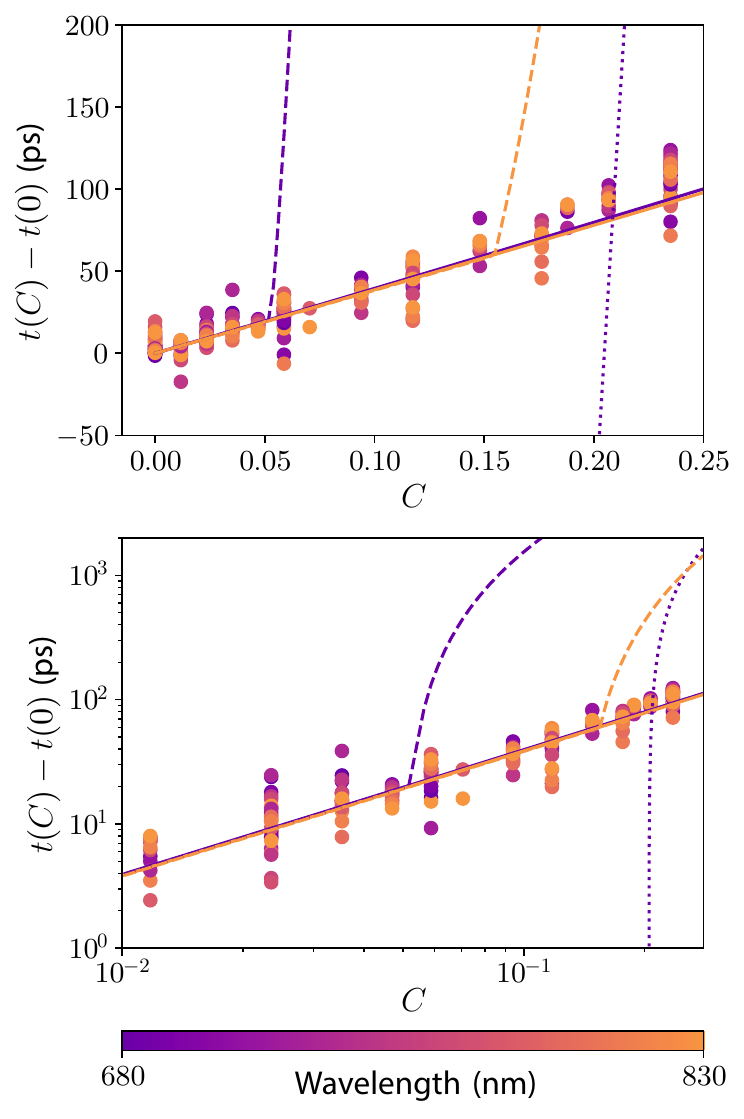}
\caption{\label{fig:alldata} EFPT differences measured experimentally, in picoseconds, versus volume concentration $C$. Both plots are the same, with different scales. Solid lines are predicted EFPTs for a photon traveling through an index-averaged medium of silica and water. Experimental values are shown as points, telegraph equation predictions are shown as dashed lines, and the diffusion approximation is shown as dotted lines. Color indicates wavelength from 680 nm (violet) to 830 nm (orange). Predictions were made using $N=5\times10^6$ photons. The orange dotted line does not appear in this figure as the values fail to cross over the measured range.}
\end{figure}

\textbf{Models for photon transport. }A classical approximation to the spatial and temporal distribution of photons is the equation of transfer ~\cite{haskell_boundary_1994,ishimaru_wave_1997}. As there is currently no general solution to  this time-dependent radiative transfer equation, most applications rely on approximations to derive a more tractable solution. Different approximation methods yield distinct solutions, with the diffusion approximation and the telegraph equation being two widely used approaches. In the following, we approximate the transmission of light through a slender tube filled with scattering medium as a one-dimensional process. We describe these two approaches below, resulting in two distinct photon probability distributions. From the probability distribution of photon locations, we derive an approximation for the extreme first passage time (EFPT) past a distance $L$. We approximate this as the first time at which the cumulative probability beyond $L$ surpasses 1/$N$, for $N$ incident photons~\cite{hass_first-passage_2024}.

The equation of transfer is given by ~\cite{haskell_boundary_1994,ishimaru_wave_1997},
\begin{align}
    \frac{1}{c} &\partial_{t} L \left(\mathbf{r},\hat{s},t\right) + \mathbf{\nabla} \cdot L\left(\mathbf{r},\hat{s},t\right)\hat{s} = -\left(\mu_{s} + \mu_{a}\right) L\left(\mathbf{r},\hat{s},t\right) \nonumber \\
    &+ \mu_{s} \int \int_{4\pi} L\left(\mathbf{r},\hat{s'},t\right) f\left(\hat{s} \cdot \hat s'\right) d\Omega' + S\left(\mathbf{r},\hat{s},t\right). \label{eq:RTE}
\end{align}
This equation determines the rate of change of the radiance, $L$, through position $\mathbf{r}$, in direction $\hat s$, and at time $t$. The radiance originates at a source, described by $S$, and is lost to the absorption coefficient, $\mu_{a}$, and scattered by the scattering coefficient, $\mu_{s}$. The speed of light in the medium is $c$, and $f(\hat s \cdot \hat s')$ is the normalized differential scattering probability for photons traveling in a direction $\hat s'$ to scatter into the direction $\hat s$. 

\eqref{eq:RTE} can be integrated over all solid angles to find a continuity equation governed by the photon density, $\varphi\left(\mathbf{r}, t\right)$, and photon flux, $\mathbf{j}\left(\mathbf{r}, t\right)$~\cite{haskell_boundary_1994}. When considering scattering in a long tube it is useful to approximate the system as quasi-one-dimensional.  Because $f\left(\hat{s}\cdot\hat{s}'\right)$ can be thought of as a distribution of scattering angles where $\cos \theta = \hat s \cdot \hat s'$, it will reduce to a single parameter, $\langle\cos{\theta}\rangle$, when integrated over all angles. For Rayleigh scattering, which is isotropic in both forward and backward directions~\cite{bohren_absorption_1983}, $\langle\cos{\theta}\rangle = 0$.


\textit{Diffusion approximation.} When the total volume concentration of scatterers greatly exceeds 1\%, \eqref{eq:RTE} can be estimated by a diffusion approximation. Under the assumption that photons interact with many particles, we assume a nearly uniform angular scattering distribution~\cite{ishimaru_wave_1997,bohren_absorption_1983}. By expressing the radiance $L\left(\mathbf{r},t\right)$ as the sum of an isotropic photon density $\varphi\left(\mathbf{r}, t\right)$ and a small directional flux, neglecting variations and anisotropy in the source term $S\left(\mathbf{r},t\right)$, integrating over all solid angles, and approximating as a one-dimensional system, we obtain~\cite{haskell_boundary_1994}
\begin{equation}\label{eq:diff}
    D \partial_{r}^{2} \varphi \left(r,t\right) - \mu_{a} c \varphi \left(r,t\right) = \partial_{t} \varphi \left(r,t\right) - c S \left(r,t\right),
\end{equation}
with diffusion coefficient
\begin{equation}\label{eq:diffD}
    D = \frac{c}{3\left(\mu_{s} + \mu_{a}\right)}.
\end{equation}
The solution to \eqref{eq:diff} is well-established as a Gaussian multiplied by a decaying exponential term that accounts for absorption~\cite{haskell_boundary_1994}.

Under this diffusion approximation photons move as Brownian walkers in a semi-infinite medium, starting at the origin and whose position we denote as $B(t)$. When absorption is minimal, the probability of a photon to be found beyond a distance $L$ at time $t$ is~\cite{redner_guide_2001,lawley_distribution_2020}
\begin{align}\label{eq:diffProb}
     P\left(B\left(t\right) \geq L\right) = 1-\textrm{erf}\left(\frac{L}{2 \sqrt{D t}}\right).
\end{align}
We approximate the expected value of the EFPT as the time at which this probability surpasses $1/N$. We note that at short times the diffusion approximation yields non-physical behavior because the maximum photon speed in this approximation is unbounded.


\textit{Telegraph equation.} The telegraph equation  offers a more accurate model for photon transport, incorporating the ballistic motion of photons. By using asymptotic expansions to approximate a solution to \eqref{eq:RTE} ~\cite{heizler_asymptotic_2012, hoenders_telegraphers_2005, gombosi_telegraph_1993}, accounting for the movement of photons in and out of each direction as opposing ``streams''~\cite{schuster_radiation_1905,durian_two-stream_1996,lemieux_diffusing-light_1998,masoliver_solutions_1992}, or by various other methods of derivation~\cite{goldstein_diffusion_1951,dudko_photon_2005,masoliver_finite-velocity_1996,weiss_first_1984,weiss_applications_2002}, we arrive at a telegraph equation for photon transport~\cite{lemieux_diffusing-light_1998}, which governs the evolution of $\varphi$ as
\begin{align}
     \partial_{r}^{2} \varphi \left(r,t\right)  = &\frac{1}{c^{2}}\partial^{2}_{t} \varphi\left(r,t\right) + \frac{1}{c}\left(2 \mu_{a} + 3 \mu_{s} \right) \partial_{t} \varphi\left(r,t\right) \nonumber \\
    &+ \mu_{a}\left(\mu_{a} + 3 \mu_{s} \right) \varphi. \label{eq:tel}
\end{align}
At short times, \eqref{eq:tel} reduces to the wave equation. At long times and in the limit of $\mu_{a}$ going to zero we recover \eqref{eq:diff}, the standard diffusion approximation, with diffusion coefficient~\cite{lemieux_diffusing-light_1998}
\begin{equation}\label{eq:telD}
    D = \frac{c}{3\mu_{s}}.
\end{equation}

The general solution to this telegraph equation is given by~\cite{goldstein_diffusion_1951,durian_photon_1997}
\begin{align}
    \varphi \left(r,t\right) &= \frac{1}{2} e^{-\left(\mu_{a}c + \gamma \right)t} \bigg[\delta \left(c t - r\right) + \delta\left(ct + r\right) \notag\\
    &+\left. \Theta\left(c t - |r|\right) \left(\frac{\gamma}{c}I_{0}\left(\frac{\gamma u}{c}\right) + \frac{\gamma t}{u}I_{1}\left(\frac{\gamma u}{c}\right)\right)\right]
\end{align}
where $\delta$ is the Dirac delta function, $\Theta$ is the Heaviside step function, $I_{0}$ and $I_{1}$ are modified Bessel functions of the first kind, $u=\sqrt{c^{2}t^{2}-r^{2}}$ represents the position relative to that of a ballistic photon, and $\gamma = \frac{3}{2} c \mu_{s}$ is a characteristic time scale for scattering events~\cite{goldstein_diffusion_1951,masoliver_solution_1993,masoliver_telegraphers_1994,masoliver_finite-velocity_1996}. We solve this numerically to obtain EFPT predictions.


\textbf{Results and discussion. }Figure \ref{fig:alldata} shows $t(C)$, the time of the peak of the EFPT distribution for scatterer volume concentration $C$, for a range of wavelengths and volume concentrations. For each wavelength measured, this time is shown relative to $t(0)$, the time of the peak of the EFPT distribution for pure water recorded at that wavelength. The measured EFPTs show no wavelength dependence and increase linearly with $C$, reaching approximately 100 ps for $C$ = 0.235. The measured EFPT differences are consistent with the extreme photons traveling at the speed of light through a \textit{non-scattering} index-averaged medium at each concentration. 

The predictions of the two models are plotted as dotted lines (diffusion approximation) and dashed lines (telegraph equation) for wavelengths at the extreme ends of our experimental range. These predictions are calculated using $N \simeq 5 \times 10^{6}$. Because the EFPT increases with decreasing $N$~\cite{lawley_distribution_2020}, this upper limit for $N$ provides a lower bound on the predicted values of the peak EFPT. Additionally, the relative difference between the predictions at 680 nm and 830 nm becomes larger as $N$ decreases. The diffusion approximation predicts non-physical results for small concentrations, allowing photons to travel faster than the speed of light for $C$ below approximately 0.21 at 680 nm and at all concentrations for 830 nm.  At higher concentrations, the diffusion approximation predicts increasing EFPT differences  for 680 nm, exceeding 650 ps (680 nm) at $C$ = 0.235. Relatively speaking, the telegraph equation does a better job and predicts a linear relationship for EFPT delay at low concentrations, determined by the speed of light in the index-averaged medium. However, for concentrations beyond approximately 0.05 (680 nm) and 0.15 (830 nm), scattering and absorption cause the EFPT difference to increase rapidly and deviate from the measured values. Maximum predicted delays of 8,000 ps (680 nm) and 900 ps (830 nm) occur at $C$ = 0.235.


\textbf{Conclusion. }The diffusion and telegraph approximations fail to accurately describe EFPT photons, either violating causality (diffusion) or predicting unrealistically large EFPT differences (diffusion and telegraph). In order to provide accurate predictions, the telegraph equation  would require either unphysically large photon counts of $N \gtrapprox 10^{16}$, far exceeding the laser output to achieve wavelength-independent results, or an anisotropy factor ($\langle \cos {\theta} \rangle \geq 0.4$) that contradicts the isotropic-scattering behavior of Rayleigh scattering. While these models are effective at predicting the behavior of the average photon in scattering media and can be used to infer bulk properties, they fail to describe extreme photons. These results suggest that assumptions made in the underlying random walk approach are incorrect. A model based on random walks which incorporate correlated or even quantum mechanical behavior may produce more accurate predictions and could better describe the underlying physical process. Such a model would unlock the potential for photon EFPT statistics to uncover critical insights about the diffusive environment, with broad applications across many areas of science~\cite{redner_guide_2001,weiss_applications_2002,godec_first_2016,noskowicz_average_1988,grebenkov_molecular_2021,polizzi_mean_2016,barney_first-passage-time_2017,zsurkis_first_2024,lawley_distribution_2020,lawley_slowest_2023,lawley_universal_2020,schuss_redundancy_2019,meerson_mortality_2015,chun_heterogeneous_2023}.

\begin{backmatter}
    \bmsection{Funding} W.M. Keck Foundation Science and Engineering grant on “Extreme Diffusion”.
    
    \bmsection{Acknowledgement} We thank M. Allgaier, J. Hass, D. Allcock, I. Corwin, and R. Parthasarathy for discussions.
    
    \bmsection{Disclosures} The authors declare no conflicts of interest.
    
    \bmsection{Data availability} Data is available upon reasonable request.
\end{backmatter}

\bibliography{main}

\end{document}